# Chemically-controlled self-assembly of hybrid plasmonic nanopores on graphene

*Giorgia Giovannini[1], Matteo Ardini[1], Nicolò Maccaferri[1], Xavier Zambrana-Puyalto[1], Gloria Panella[2], Francesco Angelucci[2], Rodolfo Ippoliti[2], Denis Garoli[1*], and Francesco De Angelis[1*]*

**Abstract**

Thanks to the spontaneous interaction between noble metals and biological scaffolds, nanomaterials with unique features can be achieved following relatively straightforward and cost-efficient synthetic procedures. Here, plasmonic silver nanorings are synthesized on a ring-like Peroxiredoxin (PRX) protein and used to assemble large arrays of functional nanostructures. The PRX protein drives the seeding growth of metal silver under wet reducing conditions, yielding nanorings with outer and inner diameters down to 28 and 3 nm, respectively. The obtained hybrid nanostructures can be deposited onto a solid-state membrane in order to prepare plasmonic nanopores. In particular, the interaction between graphene and PRX allows for the simple preparation of ordered arrays of plasmonic nanorings on a 2D-material membrane. This fabrication process can be finalized by drilling a nanometer scale pore in the middle of the ring. Fluorescence spectroscopic measurements have been used to demonstrate the plasmonic enhancement induced by the metallic ring. Finally, we support the experimental findings with some numerical simulations showing that the nanorings are endowed with a remarkable plasmonic field within the cavity. Our results represent a proof of concept of a fabrication process that could be suitable for nanopore-based technologies such as next-generation sequencing and single-molecule detection.

Functional nanomaterials are drawing the attention of different scientific communities due to their unique mechanical,[1] magnetic,[2] electrical[3] and optical properties,[4] which cannot be found in the corresponding bulk materials.[5,6,7] In this scenario, noble metallic nanostructures play a pivotal role in a wide range of applications such as electronics,[8] catalysis,[9] photonics,[10,11] sensing and biomedicine.[12,13] Among others, one of the most interesting phenomena related to metal-based nanostructures is their ability to couple with electromagnetic radiation and generate surface plasmons.[14] Surface plasmons enable the engineering and localization of enhanced electromagnetic fields which find application in diverse fields, ranging from photonics to biosensing.[15] Different approaches have been developed in order to pattern materials at the nanoscale. Among the standards fabrication techniques, thin-film deposition and nanolithography, exploit a "top-down" approach in which matter is sculpted and shaped until the desired final nanomaterial is achieved. This approach represents so far the main path in terms of robustness and efficiency although it is remarkably costly, both in terms of time and cost of the required facilities.[16,17] Colloidal lithography is emerging as a "bottom-up" fabrication technique, thanks to the possibility of producing patterns on large scales[18,19] even though control on the size and the shape still needs to be optimized.[20,21] An alternative "bottom-up" approach that is quickly emerging as a valuable and cost-efficient strategy is the one where small precursor molecules interact between each other forming nanostructures.[22] In this context, nature provides several examples of self-assembling nanomaterials that can be applied to nanotechnology.[23] In particular, DNA and proteins are considered groundbreaking tools as they naturally self-assemble into several multi-level architectures.[24] Proteins are being explored for supramolecular chemistry applications and inspire the design of novel artificial nanomaterials fabricated by biotechnological, chemical and computational expedients.[25] The access to ready-to-use protein assemblies with discrete structural patterns such as cages,[26] rings[27] and tubes[28] is leading to encouraging results, for example in magnetic resonance-based imaging.[29] As proteins, DNA has been used for similar purposes and DNA origami finds now interesting applications as scaffold for nanofabrication.[30] With respect to DNA, proteins are more versatile thanks to multiple anchoring sites, *e.g.* amino acids such as lysine, glutamate and cysteine, ready for functionalization by means of amine-, carboxyl- or thiol-reactive dyes, cross-linkers and other biomolecules.[31,32] It is noteworthy that in many proteins the carboxyl- and amine-terminal residues lie along the surface of rim and cavity thus providing two very independent surfaces for biofunctionalization.[33] Moreover, protein data banks provide a huge range of shapes,[34] size,[24] surfaces,[35] and chemical properties, enabling to achieve nanostructures with the desired morphology and features.[36]

In this context, ring-like proteins are likely to play a pivotal role as they possess unique features, *i.e.* natural high abundance, remarkable structural stability compared to most of biomolecules and protein surfaces amenable to different functionalization strategies.[37,38,39] Typical 2-Cys Peroxiredoxin family members, such as the peroxiredoxin I from the human parasite *Schistosoma mansoni* (PRX), are particular intriguing relatively to these aspects. Ten identical subunits (≈25 kDa) of PRX, under reducing physiological conditions, associate into five dimers which in turn undergo self-assembly into a large and stable ring-like complex (≈250 kDa) with thickness, outer and inner diameters of 4.5, 12 and 6 nm, respectively (PDB code: 3ZTL).[35] Subunits interact in such a way that the PRX ring's bottom and top surfaces are identical thus providing a double-faced appearance, while the inner and outer surfaces of the ring present distinct features that can be exploitable for different protein derivatization. [40]

Here we present a bio-assisted tailored synthesis of plasmonic silver nanorings using the ring-like protein PRX as a scaffold (Figure 1). Electron microscopy, absorbance and energy dispersion spectroscopy show that the surface amino acids of PRX can bind and arrange $Ag^+$ ions in aqueous solution to allow seeding growth of $Ag^0$ metal under wet reducing conditions into nanorings with inner and outer diameter down to 28 and 3 nm, respectively (PRX-AgNRs). We will demonstrate that these hybrid structures can be produced easily as colloidal suspension and integrated on-chip in order to prepare array of functional plasmonics nanostructures. In particular, we will use the affinity between PRX and graphene to achieve a site selective deposition of PRX-AgNRs in large arrays of plasmonic cavities. The possibility to finalize the fabrication with a nm-scale pore in the graphene layer enables the fabrication of "self-assembled" plasmonic nanopores. This architecture is characterized by significant electric field confinement within the cavity as theoretically demonstrated by numerical simulations and verified by fluorescence spectroscopy. These data stand out as a proof of concept that could fit for nanopore-based technologies such as single-molecule detection and next-generation sequencing.

## Results and Discussion

PRX-AgNRs were synthesized exploiting a 'bottom-up' approach based on wet chemistry. In summary, $Ag^+$ ions, formed by dissociation in water of the precursor $AgNO_3$, are firstly absorbed on the surface of PRX protein. Thanks to the ring-like conformation of the protein chosen, after reduction of the protein-bound metal by $NaBH_4$, PRX-AgNRs are achieved (Figure 1).

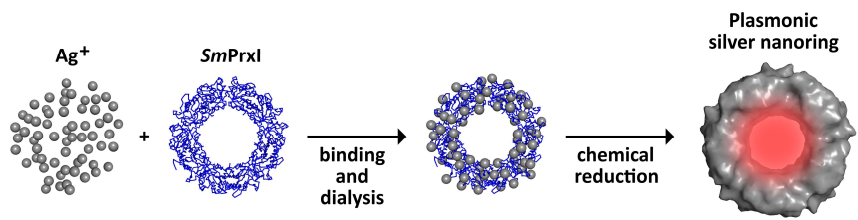

**Figure 1.** Outline of the PRX-AgNRs synthesis. The surface amino acids of the ring-like protein (PRX, shown in blue ribbons) bind and spatially arrange $Ag^+$ ions. Chemical reduction of the protein-bound $Ag^+$ ions yields metal nanorings with an average diameter of 28 nm and a cavity of 3 nm.

The PRX-AgNRs have been examined by transmission electron microscopy (TEM) as reported in Figure 2. Low magnification TEM micrographs showed the presence of electron-dense nanomaterials all over the carbon grid under different aggregation states most likely due to the presence of metal (SI – Note#1, Figure S2). When increasing the magnification, single nanoparticles with non-spherical or elliptical architecture were observed, some of which exhibited a hollow ring-like shape due to a visible inner cavity (Figure 2-A). A collection of such ring-like nanoparticles allowed the estimation of the outer diameters of the cavity to be roughly between 32 and 25 nm and the inner ones between 6 and 1.9 nm with calculated average sizes of 28±3 nm and 3.0±1.3, respectively. Micrographs clearly showed that the nanorings have an apparent rough surface suggesting that metal seeding growth occurred during the synthesis. Energy Dispersive Spectroscopy (EDS) elemental analysis confirmed that the so-achieved nanorings are effectively silver-coated PRX since the obtained color maps and related EDS spectra showed the characteristic Ag signal (main peak at 3 KeV) and other significant elements such as sulfur (S) and nitrogen (N) which are typical main components of proteins (Figure 2-B and SI, note#1 Figure S3).

The synthetic protocol was optimized in order to achieve a good yield of PRX-AgNRs and to limit the presence of undesired silver nanoparticles (AgNPs). First, considering the intrinsic reducing properties of citrate, different environmental conditions were tested for the synthesis. Dynamic light scattering (DLS) analysis proved that the reduction of $Ag^+$ and therefore the formation of AgNPs is limited using pure water instead of citrate buffer (data not showed). However, the use of water as solvent during PRX-AgNRs synthesis leads to not reproducible morphology as shown in SI - Figure S4. These results proved the importance of using citrate buffer (10 mM, pH 5.5) as solvent for the production of nicely shaped PRX-AgNRs. Being an efficient surfactant, citrate buffer was supposed to act with double role: i) citrate ions stabilize the protein in its ring-like conformation during the synthesis, an effect common to many protein complexes and ii) $Ag^+$ are better absorbed on the protein-surface likely due to the negative charge of citrate-coated protein. Due to the need of using citrate to stabilize the protein ring-like shape, the protocol has been revisited varying the ratio of $AgNO_3$:PRX used. As shown in Figure 2-C keeping constant the amount of protein used (2 µM) and varying the amount of $AgNO_3$ (in the range between 2.5-200 µM), PRX-AgNRs with the ring-like shape/morphology can be achieved in almost all cases. On the contrary, keeping constant the amount of $AgNO_3$ (100 µM) while changing the concentration of PRX led to metal PRX-AgNRs without such a nice ring-like shape (SI, note#1, Figure S5 and S6).

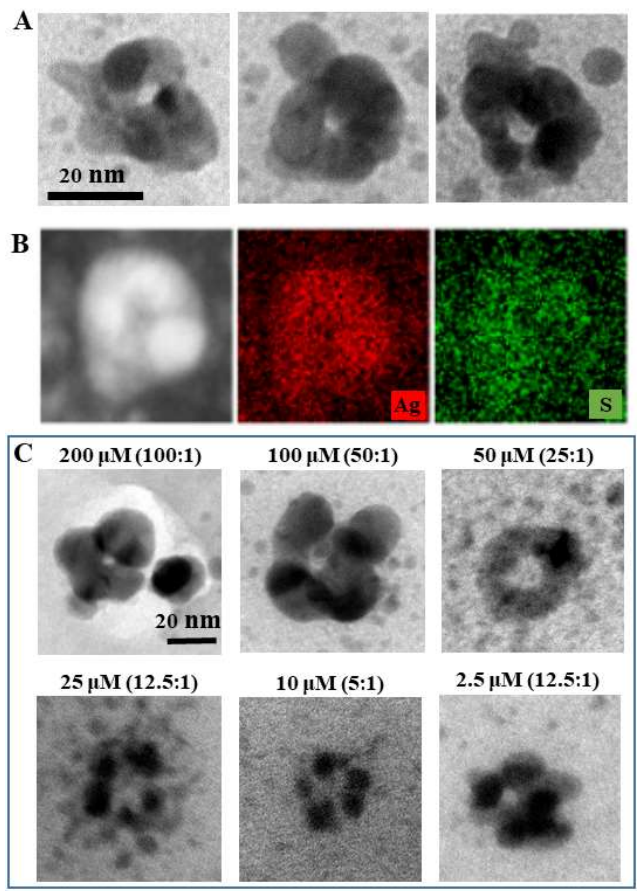

**Figure 2:** A) Ring-like nanoparticles having outer and inner diameter of 28±3 nm and 3.0±1.3 nm respectively synthesized using 100 as ratio $AgNO_3$:PRX. B) EDS analysis of Ag-PRX. Silver (red image) and sulfur (green image) are both detected for the ring-shape nanomaterial indicating its hybrid metal-bio composition. C) Example of PRX-AgNRs achieved by decreasing the amount of silver precursor used ($AgNO_3$).

As expected, the lower the $Ag^+$:PRX ratio, the thinner is the silver layer over the protein surface. Most importantly, the use of low amount $Ag^+$ allows to improve the yield of the synthesis, achieving higher amount of useful PRX-AgNRs while dramatically decreasing the amount of undesired AgNPs as noticeable in the TEM micrographs of Figure 3. The image shows that the number of PRX-AgNRs (red circles) is significantly higher compared to the undesired AgNPs (yellow circle), the latter showing very small diameters (*c.a.* 2 nm). The yield and the purity of the sample were of remarkable importance considering the final goal to be achieved with such nanostructures, that is their deposition on solid-state membrane and exploitation of their plasmonic properties. Analyzing TEM images, we approximate that using 2.5 μM of $AgNO_3$ (ratio 1.25:1 $Ag^+$:PRX) *c.a.* 80% of the nanomaterial visible on the TEM grids were identified as the wanted structures (PRX-AgNRs). This is five times higher than the percentage of ring-shape nanomaterials found using 100 μM (ratio 50:1 Ag$^+$:PRX; *c.a.* 15%). Additional details are provided in SI, note#1, Figure S7 and Table S1.

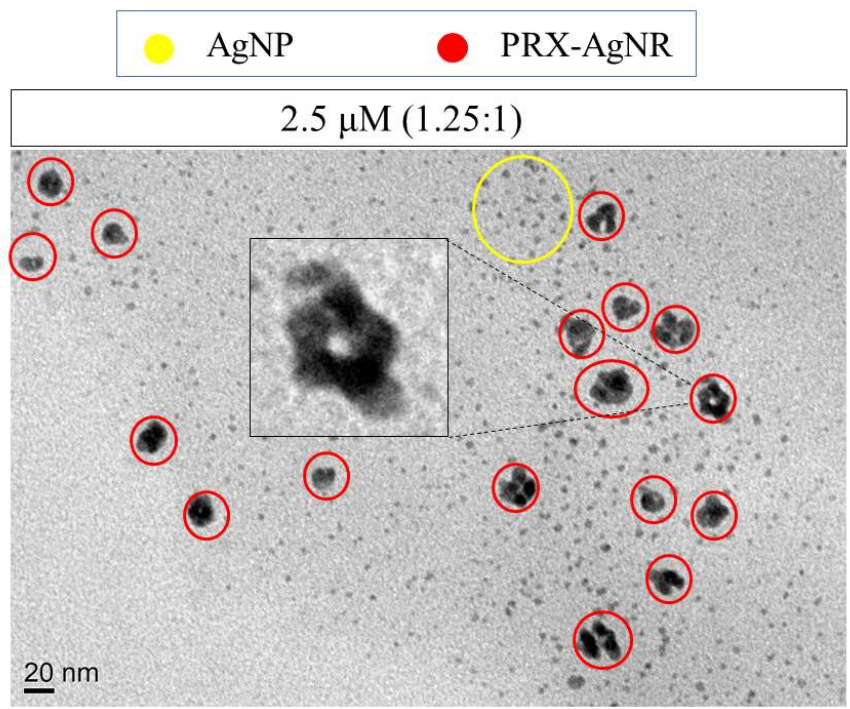

**Figure 3:** TEM image of sample prepared using 2.5 μM of AgNO$_3$. Using low amount of AgNO$_3$, PRX-AgNRs (red circles) outnumber AgNPs (yellow circle) which are very small in size (c.a. 2 nm).

Once the synthetic protocol in suspension was optimized, the attention was focused on the deposition of the PRX-AgNRs on a solid-state membrane. In principle, a simple drop casting can be used to prepare randomly distributed PRX-AgNRs as in Figure 3. Unfortunately, this is not suitable for nanotechnology applications that require instead well-organized nanostructures.[41]

In order to ensure an efficient site-selective deposition and to monitor every step of the process, *i.e.* the deposition of PRX and the successive AgNR formation/synthesis, we labelled the protein with a dye (Alexa647, suitable for fluorescent microscope analysis) and we optimized an *in-situ* synthesis of PRX-AgNRs directly on the array. In details, PRX was labelled using a PEGylated dye (PEG-Alexa647) and subsequently deposited on the membrane. While the protocol used for protein labelling is described in Material and Method section, the UV-Vis/fluorescence characterization of labelled-PRX can be found in SI note#2, Figure S8. Two alternative approaches were evaluated for the site-selective deposition of

labelled–PRX on the membrane: 1) chemical conjugation with di-thiol linker (see SI, note#3) and 2) chemical affinity with 2D materials. The better results were achieved using graphene as 2D material, as reported in Figure 4-B. The design of the final structure is illustrated in Fig. 4-A. Note that, the substrate is an array of metallic nanoholes with a graphene flakes on the bottom (details of the substrate preparation are reported in methods and in SI, note#5, Figure S11). The presence of graphene in discrete sites enables the site-selective deposition of PRX-AgNRs. Graphene, in fact, is known to promote the stable deposition of PRX rings over the surface.[42] The possibility to synthesize PRX-AgNRs with the desired morphology starting from deposited PRX was firstly tested on simple carbon and graphene TEM grids (SI, note#5, Figure S10). Once the protocol was adjusted for this particular situation, the synthetic procedure was used for the preparation of the plasmonic nanostructures directly on labelled-PRX deposited on the graphene-decorated nano-hole array. In particular, the labelled protein suspended in citrate buffer is deposited onto graphene flakes exposed in the holes of the substrate by drop casting. After all solvent evaporated, the unbound protein is removed by washing with water. A drop of $AgNO_3$ (2.5 µM) is added on the substrate and after 30 minutes removed by washing and gently drying under nitrogen. $NaBH_4$ (2.5 µM) is subsequently dropped and after 10 minute the substrate is finally washed with water. By fluorescence confocal analysis it was proved that the *in-situ* synthesis allows the localization of the nanostructure selectively in the holes of the arrays and that the fluorescence properties of the dye is not affected by the synthetic procedure (Figure 4-B). The different colors used in Fig. 4-B and 4-C want to highlight the difference in the samples (Red spots refer to labelled PRX; green spots refer to PRX-AgNRs). As noticeable from the TEM images (Figure 4-D), PRX-AgNRs synthesized directly on the array are slightly larger in comparison to the one synthesized in suspension (Figure 3). This can be explained by the favorable conditions present in correspondence of the nano-holes. Indeed, as it can be observed in the first image of Figure 4-D (inset), AgNPs tend to form at the edge of the nano-hole where artefacts of the substrate potentially induce the grown of metallic particles under mild conditions.[43] These small AgNPs can be included during the reduction of $Ag^+$ to Ag deposited on the protein surface resulting in the formation of PRX-AgNRs with larger diameter, filling all the space available in the hole. However, the inner diameter of the plasmonic ring-like nanostructure, fundamental for the flow-through applications of the structure, remains intact. In particular, we verified that a 2 nm pore can be drilled in the center of the ring, trough the graphene sheet on which the PRX-AgNRs lies, by means of well-known TEM sculpturing (see SI – Note#5, Fig. S12). Unfortunately, TEM based procedures for nanopore fabrication are expensive and time consuming. On the contrary, alternative approaches driven by the plasmonic nanostructure[44,45] can be applied to our device. In fact, a plasmonic nanostructure on an atomic thin layer of graphene has been proved to enable the self-aligned creating of pores.[45]

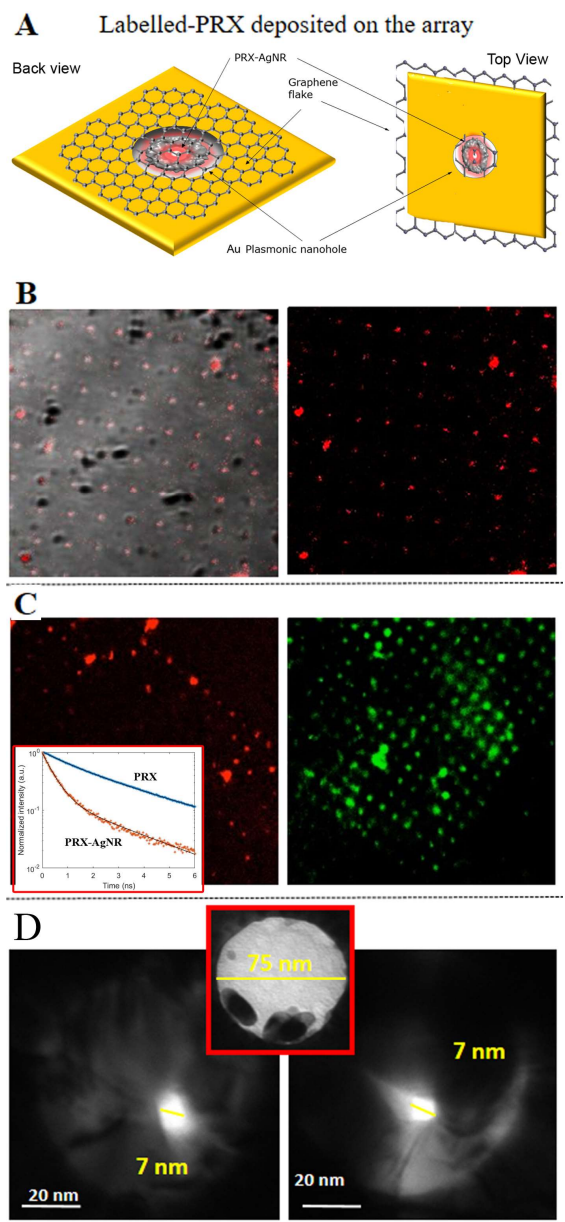

**Figure 4:** A) Cartoon of the prepared hybrid structure; B) Confocal image of labelled-PRX with PEG-Alexa647 deposited on graphene-decorated nano-hole array. C) Confocal images of respectively labelled-PRX (red rectangle) and labelled-PRX-AgNRs (green rectangle). In both cases the fluorescence is localized on the hole of the array and the fluorescence in not altered by the *in-situ* synthesis of PRX-AgNRs. Inset:

lifetime spectra of dyes on PRX and PRX-AgNRs, respectively. D) TEM images of PRX-AgNRs deposited on a single nano-hole of the array. Image with red borders shows an explicative nano-hole.

Labeling the PRX with a fluorescent dye for site-selective deposition experiments allowed us to demonstrate the method of deposition as well as to probe the potential emission fluorescence enhancement (FE) given by the plasmonic field. However, notice that two different dyes were used to acquire the images (Alexa647) and for the FE measurements (ATTORho6G). The FE properties of the PRX-AgNR structures were evaluated carrying out a comparative lifetime analysis.[46] This method is independent from the number of emitting dyes and can be used to compare PRX and PRX-AgNRs directly. Using a home-made optical set-up (see Methods), we computed the lifetime decay of the labelled PRX and the PRX-AgNR structures. First of all, two samples with graphene-decorated nano-hole arrays were fabricated. Separately, ATTORho6G labelled-PRX were deposited on the graphene-decorated nano-hole arrays. In one of the two samples, labelled-PRX-AgNRs were successfully synthesized with the *in-situ* protocol mentioned above. The lifetime of both samples was measured using our optical set-up. Different nano-holes were measured for each sample. The obtained lifetimes were (2.3 ± 0.4) ns and (1.0 ± 0.8) ns for the PRX and the PRX-AgNRs respectively. In Fig. 4B-Inset, two representative lifetime histograms are plotted. We found that the lifetime for the PRX-AgNRs is two times shorter than that of the PRX. This means that the emission of the ATTORho6G molecules is two times more efficient for the PRX-AgNRs structures. Notice that lifetime reduction, as well as local field enhancement, are a typical effect of plasmonic structures.[46] Thus, our measurements provide good evidence that plasmonic effects are present in the PRX-AgNR structures.

In order to better illustrate the expected plasmonic effects, we performed numerical simulations by means of Comsol multiphysics considering: (i) an AgNR with dimension similar to the one measured in experiments and (ii) the same AgNR deposited on a plasmonic cavity with graphene (Fig. 4-A). The results are illustrated in SI-note#6, Fig. S13. If a single AgNR is considered in $H_2O$, a significant field confinement with an intensity up to few hundreds can be achieved for an excitation wavelength around 500 nm. The AgNR suspended in $H_2O$ is not suitable for a real-application, but considering the same design deposited on a graphene layer inside a gold cavity the strong field confinement is preserved even if it is 4 times lower in intensity. Considering a nanopore also in the graphene layer (as in Figure S12), the simulation demonstrates that the field is highly confined, thus confirming the potential interest in this design.

**Conclusions**

In this paper, we optimized the synthesis of silver nanorings (PRX-AgNRs) through an easy bio-assisted method where sub 5 nm nanopores are prepared using PRX as template. The so-achieved ring-like nanostructures were selectively deposited on nano-holes arrays by exploiting the affinity of interaction between protein and graphene. As proof of concept, we tested the enhanced properties of the designed metallic nanoring: a decrease of fluorescent life-time was observed confirming the expected theoretical plasmonic effect typically founded for these kinds of nano-structures. A 2 nm pore can be drilled in the inner hole of the PRX-AgNRs deposited on the 2D support. This can enable the flow-through of molecules in the plasmonic nanoring. The plasmonic properties of PRX-AgNRs integrated on the membrane potentially allow the enhancement of signal measured from molecules flowing through silver hole. Thus, PRX-AgNRs have potential to become a promising tool for next generation sequencing or single-molecule detection as it fits with the sizes of today's forefront technologies based on biological nanopores.

**Materials and methods**

*Bacterial culture, SmPrxI (PRX) expression and purification*

PRX was obtained by heterologous expression as recombinant N-terminal 6-histidine protein according to literature[38,]. Briefly, BL21(DE3) pLysS bacteria cells (Novagen) were transformed with a PRX-encoding plasmid and grown in selective LB medium (Sigma-Aldrich) before inducing the protein expression with IPTG (Sigma-Aldrich). Cells were harvested by centrifugation and suspended in TRIS buffer pH 8.5 containing 150 mM NaCl before lysis by sonication. A clarified cell extract was obtained by ultracentrifugation and loaded onto a nickel-bound affinity column connected to an ÄKTAprime plus chromatography system (GE Healthcare). The protein was stripped out by flushing with imidazole (Sigma-Aldrich) and its purity assessed through non-native polyacrylamide gel electrophoresis (Sigma-Aldrich). The protein concentration was estimated by spectrophotometry according to a molar extinction coefficient and a ring molecular weight of 25440 $M^{-1}$ $cm^{-1}$ and 252.7 kDa, respectively. EDTA (EuroClone) and 2-ME (Sigma) were then added to remove any residual nickel and keep the protein in a ring conformation. Finally, the purified solution was sterilized by filter membranes (MDI Membrane Technologies, LLC) and stored at 4 °C up to four months. The protein concentration refers to the single subunit (25 kDa) throughout the text.

*Silver nanorings synthesis*

Firstly, the purified PRX was dialyzed in 10 mM sodium citrate buffer pH 5.5 (Sigma-Aldrich) for buffer exchange and removal of contaminants such as NaCl, EDTA and β-ME. Dialysis was performed overnight at 8 °C using 15 kDa cut-off regenerated cellulose membranes (ThermoFisher Scientific) according to a volume ratio of 2500 (citrate buffer:PRX). The dialyzed protein was centrifuged 5 min at 10000 rpm at 8 °C to remove any aggregate and its concentration estimated by spectrophotometry (Nanodrop). The protein concentration was set at the desired concentration using citrate buffer as diluting solvent. As a metal precursor, a stock solution of silver nitrate ($AgNO_3$, Sigma-Aldrich) in milliQ water was freshly made in a 1.5 mL low-adsorption tube (Sigma-Aldrich) and protected to light prior to use. Using 200 μL as final volume, 180 μL of PRX suspension was prepared at the desired protein concentration. To this solution 20 μL of $AgNO_3$ was added reaching the desired final concentration. In particular, $AgNO_3$ was added under shaking (500 rpm) at room temperature (R.T.; 20 °C) 5 μL over 10 minutes (5 μL x 4). Table below shows the concentration of $AgNO_3$ and PRX respectively used in this study and the corresponding ratio.

| Sample | $AgNO_3$ (μM) | PRX (μM) | $AgNO_3$:PRX |
|---|---|---|---|
| 1 | 200 | 2 | 100 |
| 2 | 100 | 2 | 50 |
| 3 | 50 | 2 | 25 |
| 4 | 25 | 2 | 12.5 |
| **5** | 10  | 2 | 5    |
| **6** | 5   | 2 | 2.5  |
| **7** | 2.5 | 2 | 1.25 |
| **8** | 1   | 2 | 0.5  |

Similarly, a buffered protein-free solution was made and used as control sample. Both samples were dialyzed 2 h at 8°C under orbital shaking at 150 rpm in citrate buffer according to a volume ratio of 2500 (citrate buffer:sample) to remove the excess of $Ag^+$. After dialysis, a stock solution of reducing agent sodium borohydride ($NaBH_4$, Sigma-Aldrich) was freshly prepared in milliQ water in a low-adsorption tube and it was added to the dialyzed samples under constant shaking (500 rpm) to induce chemical reduction of silver. A total volume of 20 µL of $NaBH_4$ was added dropwise (5 µLx4) over 10 min of constant shaking at R.T. starting from a stock solution with specific concentration in order to achieve the desired concentration in the reaction. The final concentration of $NaBH_4$ always corresponded to the concentration of $AgNO_3$ previously chosen. Samples were then dialysis against citrate buffer for 2 h. Note that citrate buffer was chosen as suitable diluting agent for $AgNO_3$ as other buffering reagents such as TRIS, HEPES and MOPS caused fast precipitation of the salt probably due to chelation of Ag+ ions and formation insoluble complexes (data not shown).

*Labelling of PRX and PRX-AgNR*

Depending upon the type of analysis, two different dyes have been used to labelled the biological and hybrid materials: Alexa647 651-672 nm as respectively $\lambda_{ex}$- $\lambda_{em}$ (Alexa Fluor™ 647 Succinimidyl Ester; ThermoFisher) was used for confocal imaging while ATTORho6G 533-557 nm as respectively $\lambda_{ex}$- $\lambda_{em}$ (ATTO Rho6G Succinimidyl Ester; ATTO-TEC) was used for the evaluation of fluorescence lifetime.

In both cases, the dye was solubilized in 10 mM PBS pH 8 and $NH_2$-PEG-SH, 1kDa (creative PEGWORKS) was added using as equivalent molar ration 1.5:1 (Dye:PEG). The reaction was shacked overnight at R.T.

and finally dialysis was performed overnight at 8 °C using 300-500 Da cut-off regenerated cellulose membranes (ThermoFisher Scientific) against 1 L of DI water.

10 μL of the PEGylated dye was added to the suspension of PRX reaching 100 μM as final concentration and the mixture was shacked (500 rpm) overnight at 8°C. The biological/hybrid material was then purified by dialysis against 500 mL of citrate buffer using 15 kDa cut-off membrane for a minimum of 6 hours. The so achieved labelled nanomaterials were characterized by spectrophotometric analysis (Tecan, Infinite M200).

*In-situ synthesis*

Labelled-PRX was firstly deposited on the substrate (carbon/graphene TEM grids or membranes with graphene-decorated nano-hole array): 5 μL of labelled-PRX at the desired concentration (2 μM) was dropped on the substrate. Once the solvent evaporated, the substrate was washed by immersion in MilliQ for 30 seconds and dried gently under nitrogen. Subsequently, 5 μL of $AgNO_3$ solution in citrate buffer at the desired concentration (2.5 μM) was dropped on the substrate. After 30 minutes the grid was washed again, dried with nitrogen and 5 μL of $NaBH_4$ at the same concentration as $AgNO_3$, was dropped on the surface. After 5 minutes the substrate was finally washed and dried.

*Bright-Field Transmission Electron Microscopy*

Bright-Field Transmission Electron Microscopy (BF-TEM) imaging was carried out to investigate the morphology of samples by means of a JEM 1011 microscope (JOEL) equipped with a tungsten filament operating at 80 keV. To this aim, 2.5 μL (x2) of freshly prepared samples were dropped onto double carbon films (ultrathin carbon on holey carbon) on copper grids (Agar Scientific Ltd.) and let completely dry at room temperature under chemical hood. The grids were then rinsed three times with milliQ and gently dried with a mild nitrogen stream water before imaging by BF-TEM.

*Preparation of nano-hole array and electrodeposition of graphene sheets*

The substrate used for site-selective PRX deposition (illustrated in Fig. 4) has been prepared following procedure recently reported by our group.[47] In summary, an array of nanoholes is prepared by means of FIB milling into a 100 nm thick Si$_3$N$_4$ membrane coated with 5//95 nm of Ti//Au. The diameter of the holes can be tuned by choosing the most suitable ion current, in our case holes of 60 nm have been prepared. These nanoholes have been then plugged with exfoliated graphene flakes[48] by using electrophoretic deposition.[47] An example of the prepared substrates is reported in Fig. S11. It is worth noting that the back-side of the substrate is coated with gold (see Fig. 4A), while the front side is coated with Si$_3$N$_4$. The graphene flakes are deposited on the gold layer, while the PRX was drop casted on the front side. This enables the preferential deposition of PRX on graphene with respect to Si$_3$N$_4$ where the PRX can be easily washed away.

*Confocal microscope analysis*

Labelled-PRX deposited on a graphene-decorated nano-hole array and PRX-AgNRs synthesized *in-situ* were imaged using a Nikon A1 confocal microscope. The structures were labelled with Alexa647 (651-672 nm $\lambda_{ex}$- $\lambda_{em}$). The images were taken with a laser at 640nm, and the fluorescence was collected in the 650-700nm range.

*Fluorescence lifetime analysis*

The Fluorescence lifetime analysis of both the labelled-PRX deposited on a graphene-decorated nano-hole array and the labelled PRX-AgNRs synthesized *in-situ*, have been done using a non-commercial optical set-up. An avalanche photodiode (APD), as well as the picosecond laser at 532nm are connected to a time-correlated single-photon counting module in time-resolved mode. Making use of a home-built code, we build a histogram of 300 bins, each of them having a temporal width of 30 ps. A certain time delay is applied to the laser channel, so that the histogram is monotonously decreasing. The histogram measurement is carried out for 150 s. A biexponential function of the kind $Ae^{-t/\tau_A} + Be^{-t/\tau_B}$ is used to fit the decay curve. We estimate the lifetime as $\tau = \frac{(A\tau_A + B\tau_B)}{A+B}$. Each measurement is repeated several times, and the result is given as the average plus an uncertainty given by the standard deviation. More information on the optical set-up can be looked up in our previous publication.[46] ATTORho6G dyes (533-557 nm $\lambda_{ex}$- $\lambda_{em}$) were used for the labelling.

*Energy-Dispersive X-ray Spectroscopy (EDS)*

Elemental analysis was carried out by Energy-dispersive X-ray spectroscopy (EDS) by means of a Bruker XFlash5060 SDD system installed on the same TEM.

*Numerical simulations*

Numerical simulations were carried out to investigate the optical response of the PRX-AgNRs. The electromagnetic response of an isolated nanoring was simulated using the Finite-Element Method (FEM) implemented in the RF Module of Comsol Multiphysics®. The dimensions of the outer and inner diameters of the rings were set according to the average sizes obtained during BF-TEM investigations. The model computes the electric field of the nanoring. The unit cell was set to be 300 nm wide in both x-, y- and z-directions, with perfect matching layers (200 nm thick) at the borders. An unpolarized plane wave impinges on the structure.

*UV-Visible spectroscopy*

UV-Vis analysis of the formation of PRX-AgNRs in solution was analysed recording the absorption scan (230-800 nm range) of the sample at each step of the synthesis. The synthesis was carried on in a quartz cuvette (Hellma) and the measurements were carried on with a Cary300 (Agilent).

# SUPPORTING INFORMATION

**Supporting Note #1: Synthetic protocol**

*UV-Vis analysis*

The extinction spectra in the UV-Visible wavelengths range 235-800 nm of PRX and AgNO$_3$ in citrate buffer before and after mixing, dialysis and treatment with NaBH$_4$ are shown in Figure S1. The spectrum of pure buffer is also reported as control showing no significant optical features in the whole range. Likewise, AgNO$_3$ was almost transparent to light. Conversely, PRX exhibited the typical spectrum in the UV range, that is a first sharpen band between 260 and 290 nm belonging to the aromatic residues and a second one below 240 nm related to peptide bond. Accordingly, no further features in the visible region were detected due to the absence of chromophores as reported.[1] As the NaBH$_4$ reducing agent was added to the PRX-AgNO$_3$ sample, the color turned from colorless to very pale yellow (image not shown) and a new band was detected within the range 350-700 nm. Importantly, no precipitation was observed meaning that the solution remained colloidal. The 350-700 nm band was reasonably ascribed to SPR of metal silver Ag$^0$ formed upon reduction of Ag$^+$ ions whose surface conduction electrons are known to strongly interact with light and undergo to collective oscillation.[2] Indeed, the SPR between 400 (violet light) and 500 nm (green light) is typical of citrate-stabilized metal silver nanoparticles with spherical shape and can be tuned by changing the particle size.[3] Therefore, broadening of SPR up to 700 nm suggested that other nanostructures likely as aggregates and non-spherical nanoparticles formed during the process. Indeed, non-spherical silver nanoparticles exhibit red shift of the SPR to wavelengths above 500 nm.[4] Thus, it is admitted that a mix of shaped nanoparticles and aggregates contributed to the SPR extinction band ascribed to metal silver. Similarly, red shifting is observed when nanoparticles undergo aggregation that makes the conduction electrons delocalized and shared amongst neighboring particles.[5] On the other hand, the protein-free sample did not exhibit any optical activity after addition of NaBH$_4$ due to the absence of AgNO$_3$ removed by dialysis (data not shown). Taken together, these data suggested that the presence of the protein did not prevent the reduction process of silver from ion to metal form while however affecting its nucleation during the nanoparticle synthesis.

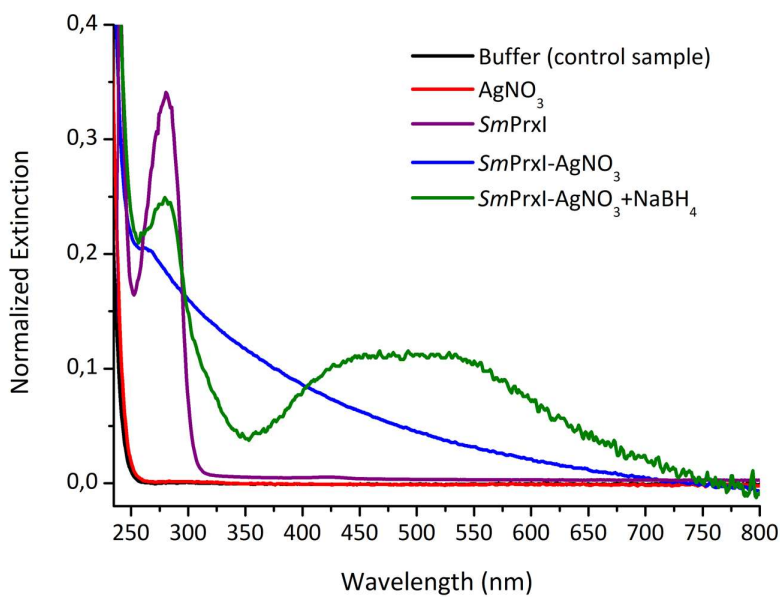

**Figure S1:** UV-Vis extinction spectra of PRX and AgNO$_3$ in citrate buffer solution before and after mixing, dialysis and treatment with NaBH$_4$. Both the buffer and AgNO$_3$ were transparent to light between 235 and 800 nm. Conversely, PRX showed the typical extinction bands between 260 and 290 nm and below 240 nm. After mixing them up, the protein's extinction is broadened above 300 nm likely due to aggregation of molecules induced by binding of Ag$^+$. Upon addition of NaBH$_4$ to the PRX-AgNO$_3$ solution, a new broad band between 350 and 700 nm appeared reasonably due to growing of metal silver nanoparticles and aggregates.

*Morphological and chemical properties*

The PRX-AgNO$_3$ sample after treatment with NaBH$_4$ have been examined by BF-TEM (Figure S2). Low magnification BF-TEM micrographs showed the presence of electron-dense nanomaterials all over the carbon grid under different aggregation states most likely due to the presence of metal (Figure S2-a), therefore supporting the extinction band 350-700 nm of metal silver observed by optical measurements (Figure S1, green spectrum). When increasing the magnification, single nanoparticles with non-spherical or elliptical architecture were observed, some of which exhibiting a hollow ring-like shape due to a visible inner cavity (Figure S 2-b). A collection of such ring-like nanoparticles allowed the estimation of the outer diameters roughly between 32 and 25 nm and the inner ones of to the cavity between 6 and 1.9 nm with calculated average sizes of 28±2.6 nm and 3±1.27, respectively (Figure S2-c). Micrographs clearly showed that nanorings have an apparent rough surface thus suggesting metal seeding growth. According to the size of the PRX ring estimated though X-ray crystallography, *i.e.* 12 and 6 nm,[1] the so-obtained nanorings were ascribed to seeding growth of a metal silver shell onto the surface of the PRX ring scaffold thus causing thickening of the protein ring.

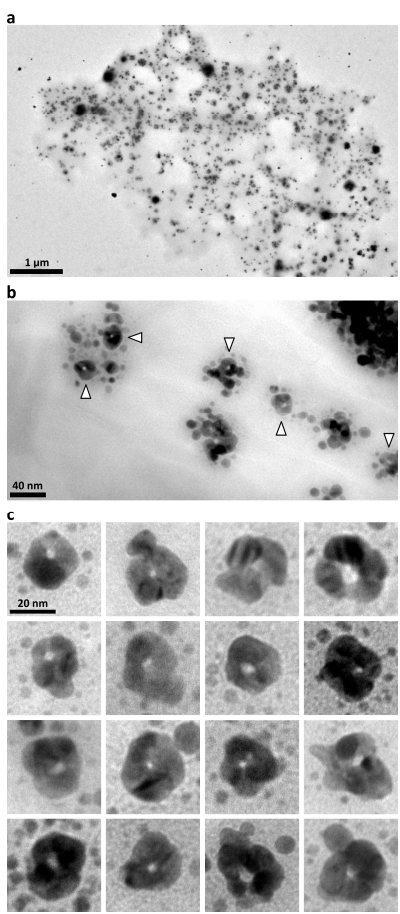

**Figure S2:** Morphological structure of the PRX-AgNO$_3$ sample after treatment with NaBH$_4$. a) Low-magnification BF-TEM micrographs showing electron-dense particles likely due to metal silver under several aggregation states. b) High-magnification micrographs where single hollow non-spherical nanoparticles (highlighted by arrowheads) are detected exhibiting a ring-like shape due to a visible cavity. c) Collection of ring-like nanoparticles, which exhibit outer and inner diameters of 28±2.6 nm and 3±1.27, respectively. The nanorings' shape and sizes suggested that seeding growth of metal silver occurred under reducing conditions onto the surface of the PRX scaffold.

*EDS elemental analysis*

EDS elemental analysis were carried out at -180 °C to avoid carbon contamination induced by focused electron beam, which dominates at room temperature (data not shown). The resulting color maps and related EDS spectrum showed the signal characteristics of elemental silver with the main peak at 3 KeV due to the SPR property. Importantly, other significant elements were found such as sulfur and nitrogen reasonably coming from the main protein's components, *i.e.* amino acids (Figure S3-a). Additional elements were also detected, *i.e.* sodium and chloride, and ascribed to residual NaCl salt contaminants belonging to buffer solution (TRIS-NaCl) used throughout the protein purification step (see Materials and Methods). Similar results were obtained when analyzing those nanorings whose cavity was only barely visible revealing the signals of both silver and protein (Figure S3-b). On the other, only silver was detected when analyzing nanoparticles with no cavity and apparent smooth spherical shape (Figure S3-c).

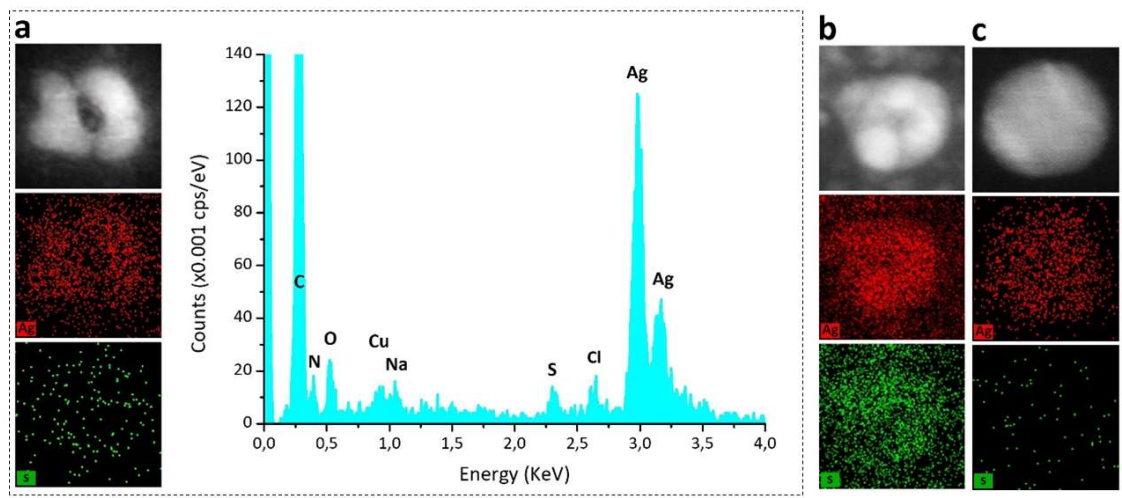

**Figure S3:** a) and b) results of EDS elemental analysis on nanorings in which silver (Ag) was detected along with other elements such as sulfur (S) and nitrogen (N) which are related to the biological material. c) Elemental analysis of simple silver nanoparticle. In this case, only silver was detected whereas no elements related to the protein.

The synthetic protocol was optimized in order to achieve a good yield of PRX-AgNRs and reduce the presence of undesired AgNPs in the final sample. Considering the reducing properties of citrate buffer, the synthesis was accomplished in MilliQ water. In these conditions the spontaneous formation of AuNPs is limited as confirmed by DLS studies (data not shown). However, as clearly visible in Figure S 4, the morphology of the obtained PRX-AgNRs do not fit with the requirement independently to the ratio AgNO$_3$:PRX chosen. In particular, the ring shape is completely or partially lost. This could be due to the change in conformation of the protein in such environmental conditions.

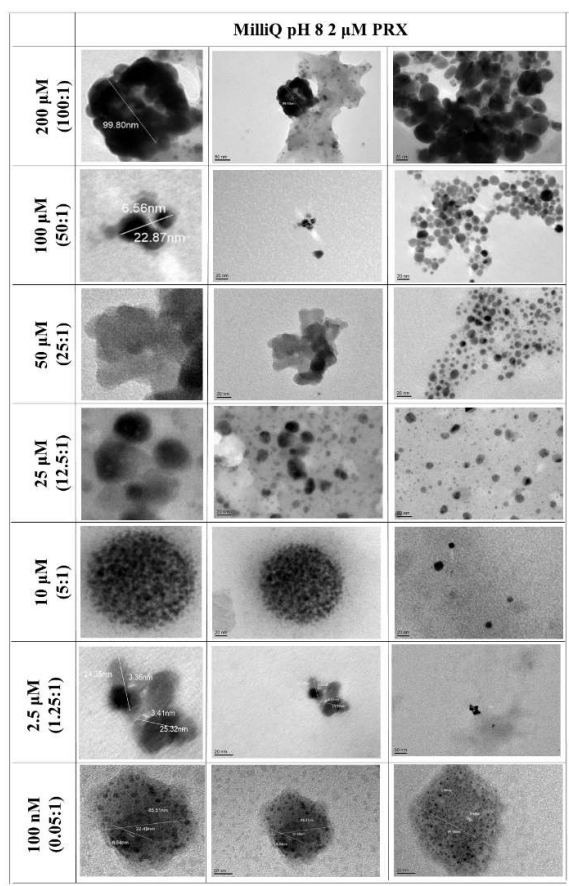

**Figure S4:** TEM images showing the morphology of PRX-AgNRs obtained using MilliQ water instead of citrate buffer. The ring shape is completely/partially lost independently to the ratio AgNO$_3$:PRX used.

Figure S 5 shows some examples of PRX-AgNRs synthesized using citrate buffer 10 mM pH 5.5 as solvent but varying the ration between AgNO$_3$:PRX. In particular, the concentration of PRX was kept constant (2 µM) whereas the concentration of AgNO$_3$ was decreased (200-1 µM). As noticeable, the layer of silver becomes thinner decreasing the concentration of precursor (AgNP$_3$) but the ring shape is maintained, that is fundamental for their application in flow through detection approach.

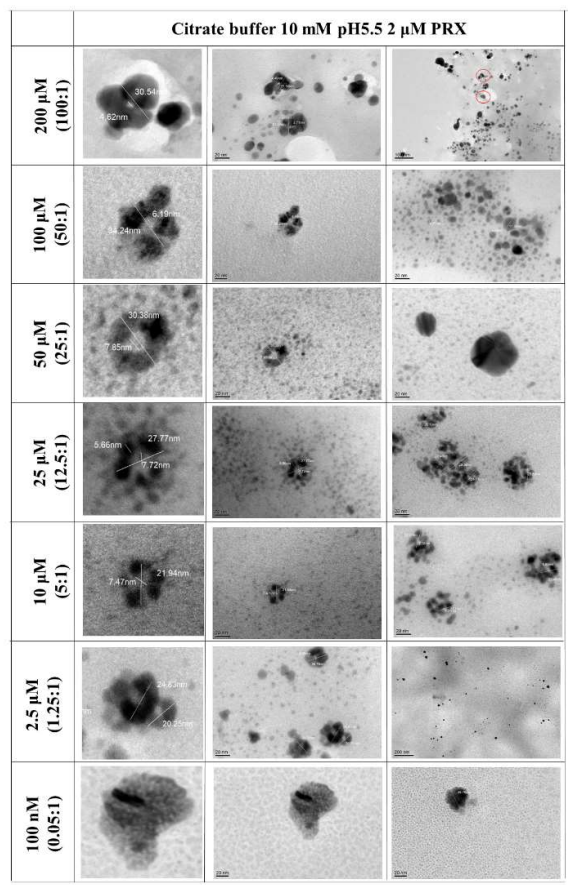

**Figure S5:** TEM images of PRX-AgNRs obtained using citrate buffer as solvent and decreasing the ratio between AgNO$_3$:PRX. Even though the silver layer becomes slightly thinner lower is the ratio used, the metallic nanomaterial maintains the ring shape that is fundamental for the designed application: flow through single molecular detection or protein/DNA sequencing.

Worth of noticed, different results were observed changing the amount of PRX and keeping constant the amount of silver precursor. Indeed, even though the ratio AgNO$_3$:PRX was the same to the one shown above, ring-like PRX-AgNRs were obtained only when 2 µM was used as PRX concentration as shown in Figure S6. This proves the delicate equilibrium required for the synthesis of such particular hybrid nanomaterials.

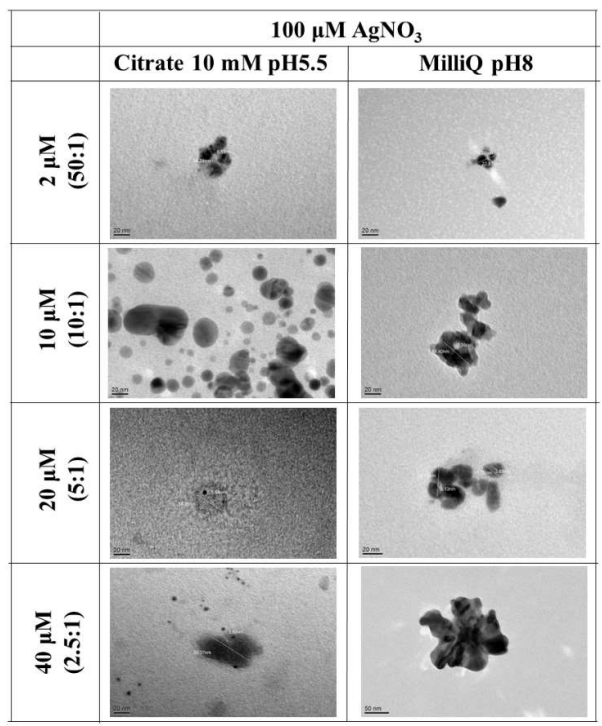

**Figure S6:** TEM images of PRX-AgNRs obtained in citrate and MilliQ water. In this case, the amount of PRX was varied keeping constant the amount of silver precursor. Even though the final ratio is comparable to the once shown previously, in such conditions the final nanomaterials have not always the ring-shape needed.

Using different ratio between $AgNO_3$:PRX it was possible to optimize the yield of the synthesis and the purity of the final sample. As shown in Figure S7, decreasing the amount of silver precursor the amount of undesired AgNPs (yellow circles) decreases whereas an increase in the number of PRX-AgNRs can be observed (red circles).

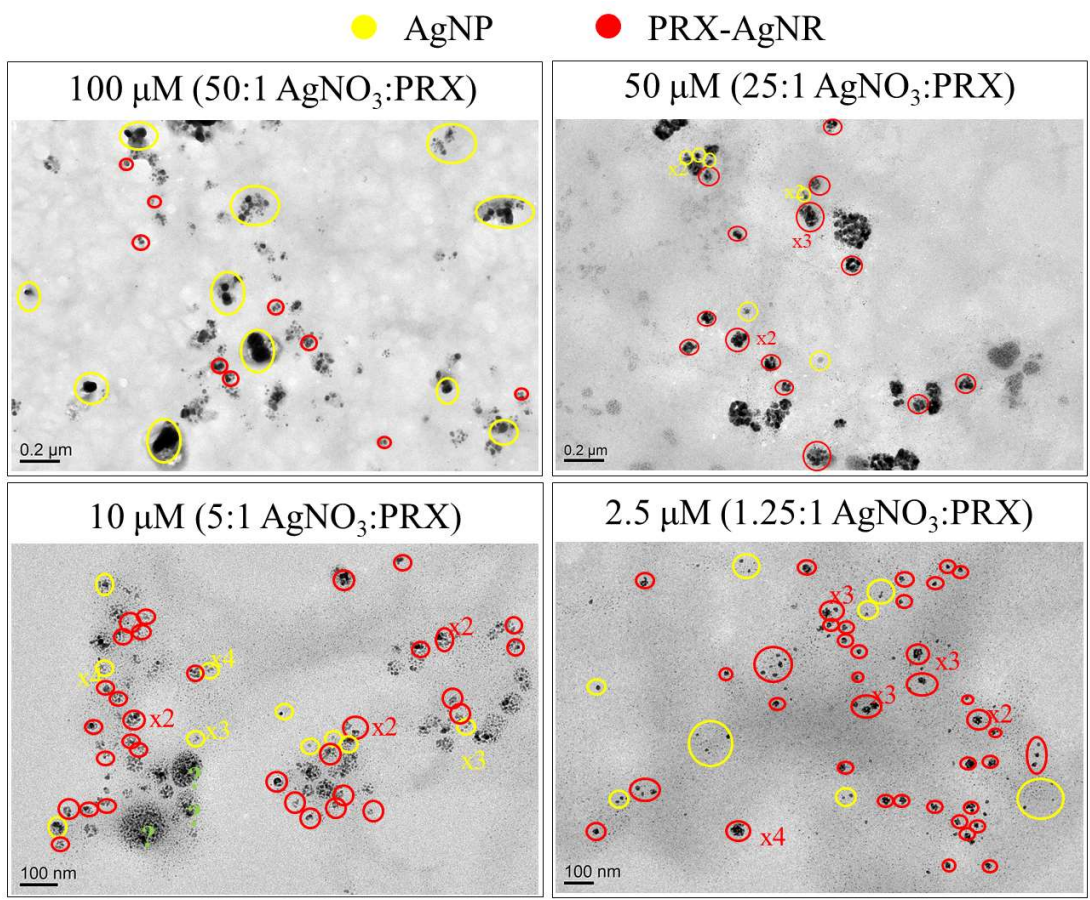

**Figure S7:** Example of TEM images sowing the improvement in yield and purity of the synthesis after optimization of the protocol. The yield was calculated counting respectively AgNP and PRX-AgNR in 4 mashes of TEM grid for each sample.

In Table S1 are reported the yield calculated referred to samples with 100, 50, 10 and 2.5 μM of AgNO$_3$. The values, found counting AuNP and PRX-AuNR deposited on TEM grids, confirmed that decreasing the amount of Ag$^+$ used for the synthesis the amount of undesired AgNP synthesised decreases as well while/consequently increases the yield of the wanted PRX-AgNRs.

**Table S1:** Yield of AgNPs and AgNRs obtained reacting a constant amount of PRX (2 μM) with 100, 50, 10 and 2.5 μM of AgNP$_3$. Values are determined counting AuNP and AgNR present in 4 mashes of TEM grid.

|  | AgNP | PRX-AgNR |
|---|---|---|
| **100 μM (50:1)** | 86 % | 14 % |
| **50 µM (25:1)** | 63 % | 37 % |
| **10 µM (5:1)** | 53 % | 47 % |
| **2.5 µM (1.25:1)** | 19 % | 81 % |

**Supporting Note #2: Labelling of PRX and PRX-AgNR**

After dialysis, the fluorescence properties of labelled PRX and PRX-AgNRs were defined by spectrophotometric analysis (Figure S8). In this example were considered PRX-AgNRs samples prepared with 50, 10 and 2.5 µM of $AgNO_3$. Figure S8-A and B show respectively the absorbance and fluorescent spectrum of each sample. The signal measured for PRX (violet line) is lower compared to PRX-AgNRs samples (scale of green lines). In this case, the difference it is likely related to the mount of dye attached on the nanomaterial since at this excitation (Alexa647) no fluorescent enhancement is expected. Using the signal measured for the dye solution at the concentration used for the synthesis as reference, the amount of Alexa647 (Figure S8-C) and the yield of the functionalisation (Figure S8-D) were calculated.

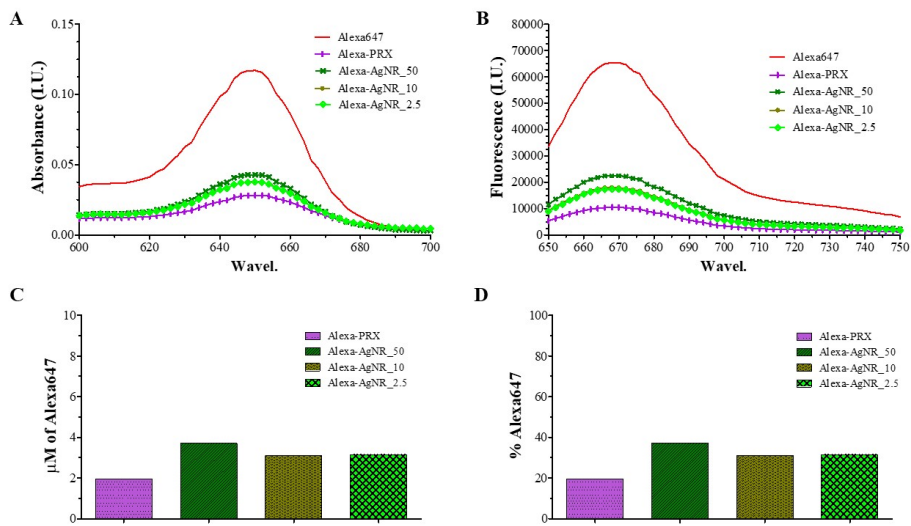

**Figure S8:** Characterization of labelled PRX and labelled PRX-AgNRs. A) and B) show respectively absorbance and fluorescent spectrums of each sample. C) Calculated concentration (µM) of dye present on PRX (violet bar) and on PRX-AgNRs synthesized using different ratio (green scale bars). D) Calculated yield (%) of the functionalization procedure.

**Supporting Note #3: Deposition protocol**

Different protocols were tested in order to achieve an efficient deposition of PRX in the nano-holes of the arrays. In case 1 and 2, flakes of respectively graphene and $WS_2$ were firstly deposited by electrophoresis on one side (gold side) of the membrane; on the opposite site ($Si_3N_4$ side) labelled-PRX was deposited by drop-casting. In case 3, the gold side of the membrane was functionalized with a dithiol linker (1, 9-Dimercaptononane). Labelled-PRX was added on the other side where the presence of SH groups exposed inside the nano-holes should ensure the selective decoration of the array. In Figure S9 are shown the confocal images of the results of deposition. As noticeable, method 1 gave better results in terms of efficiency and efficacy. Likely due to the affinity of protein for graphene, fluorescence (therefore labelled-PRX) is localized basically only inside the holes of the array. In the other cases, the fluorescence is not only localized in correspondence of the holes (case 2, $WS_2$ flakes) or the efficiency of the deposition is low (case 3, dithiol linker).

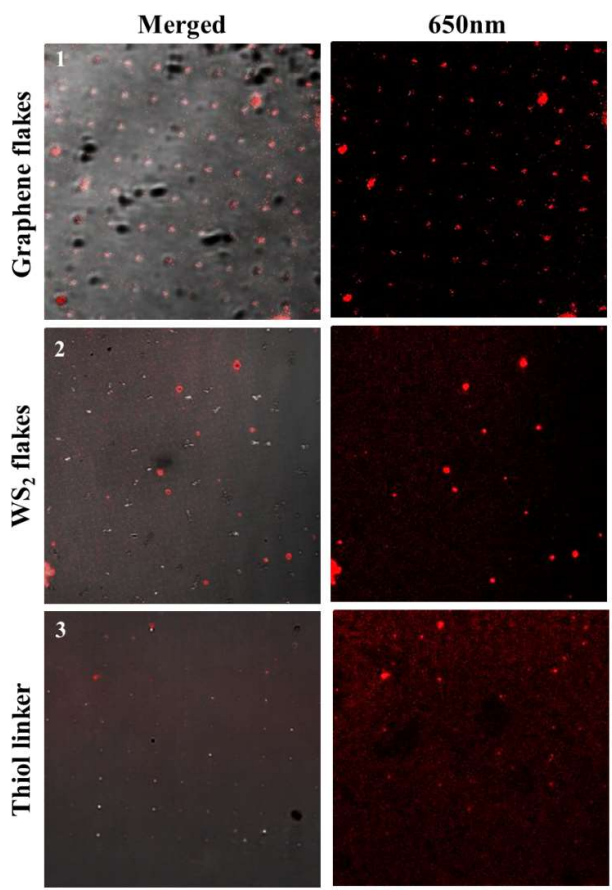

**Figure S9:** Confocal images of results achieved with the three deposition methods of labelled-PRX tested.

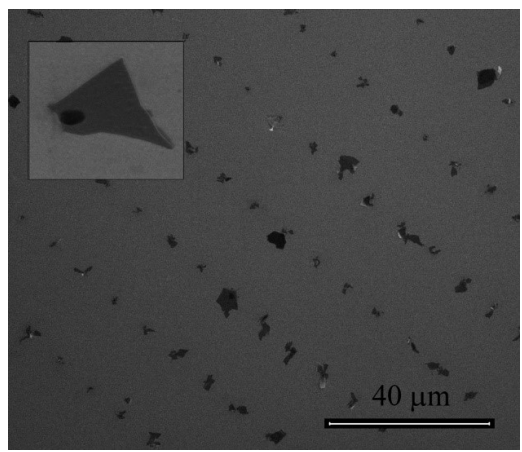

**Figure S10:** Example of site-selective deposition of graphene flakes on gold plasmonic nanoholes.[6]

**Supporting Note #5:** *In-situ* synthesis of PRX-AgNR

The protocol for an efficient *in-situ* synthesis of PRX-AgNRs was optimized firstly on carbon and graphene TEM grids. In both cases, ring-shape nanomaterial were found on the grid.

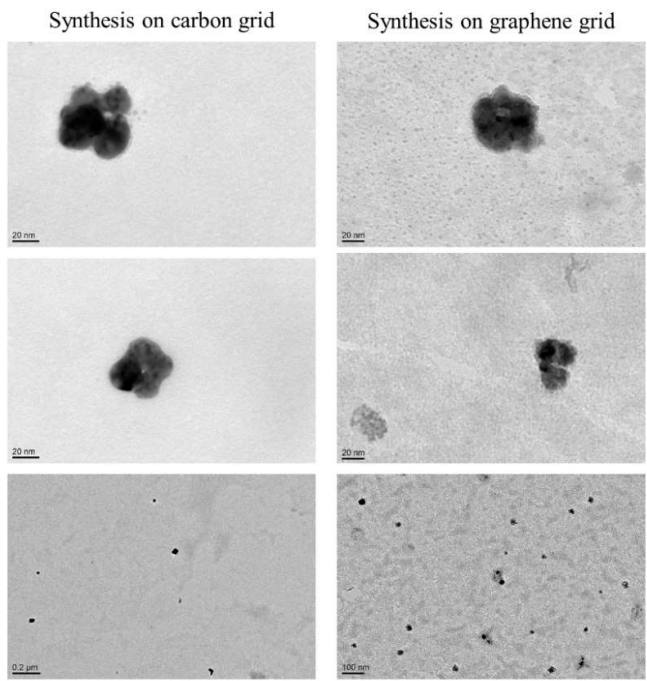

**Figure S10~~11~~:** PRX-AgNR synthesised *in-situ* on carbon (left) and graphene (right) TEM grids.

**Supporting Note #5: TEM sculpturing of nanopore on PRX-AgNR-Graphene**

For drilling experiments, the samples were prepared on double carbon films (ultrathin carbon on holey carbon) on Cu TEM grids. Electron beam drilling was performed within an image-corrected JEOL JEM-2200FS TEM (Schottky emitter), operated at an acceleration voltage of 200 kV. For these experiments the specimens were loaded into a cryotransfer holder (Gatan, model 626), cooled down and kept at the lowest temperature (about -180° C) throughout the experiment. This proved to be a successful strategy to prevent carbon contamination, otherwise induced by the focused electron beam irradiation needed. Drilling was performed in STEM mode by using a 2 nm spot (the largest available), with a convergence angle of 17 mrad, at high magnification (in the 106 range). The optimum irradiation time for hole formation in these conditions was 20 seconds, as demonstrated by HAADF-STEM imaging of the areas of interest.

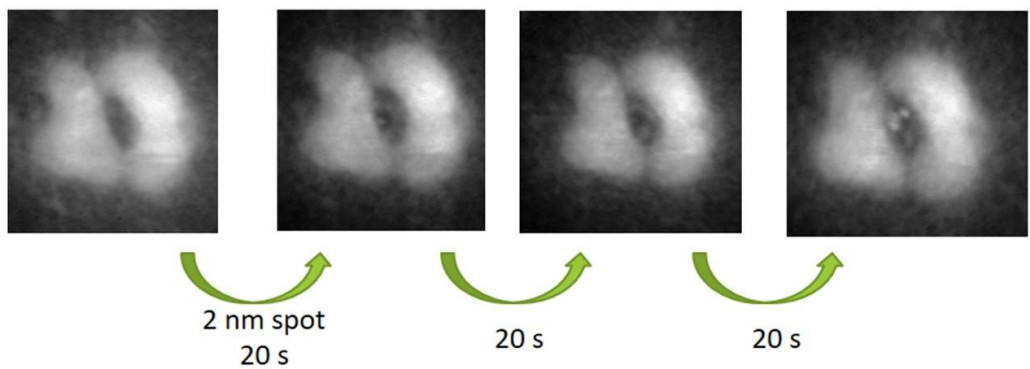

**Figure S11<del>1</del>2:** PRX-AgNR on graphene. Successive steps of TEM sculpturing of a nanopore.

**Supporting Note #6: Plasmonic properties simulation**

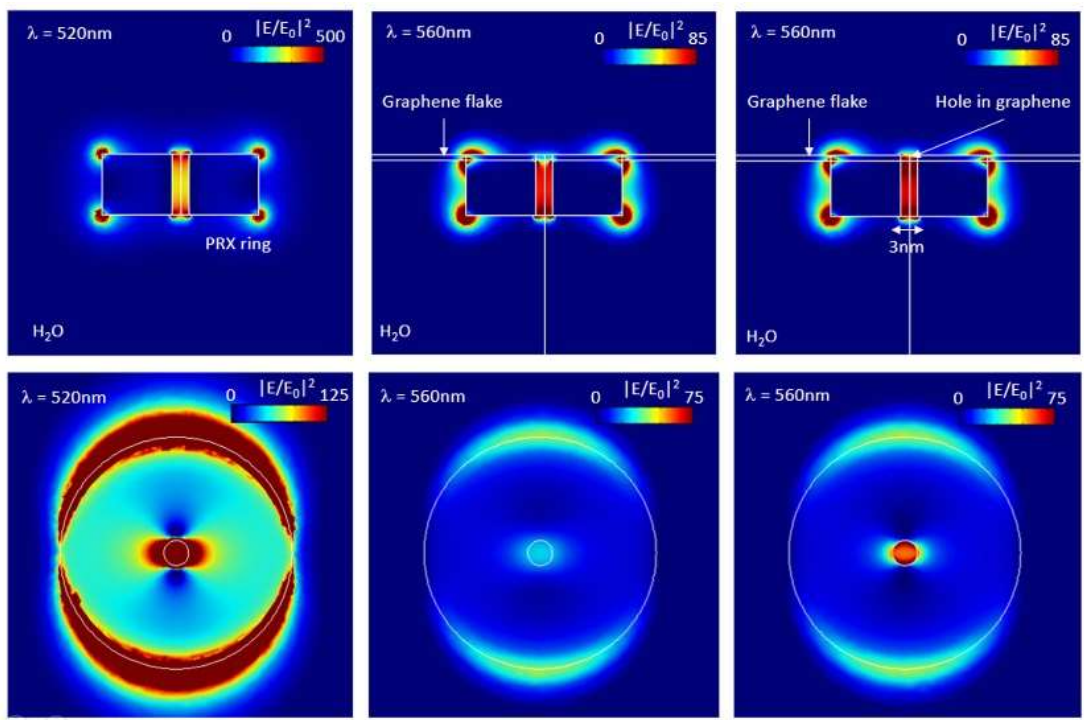

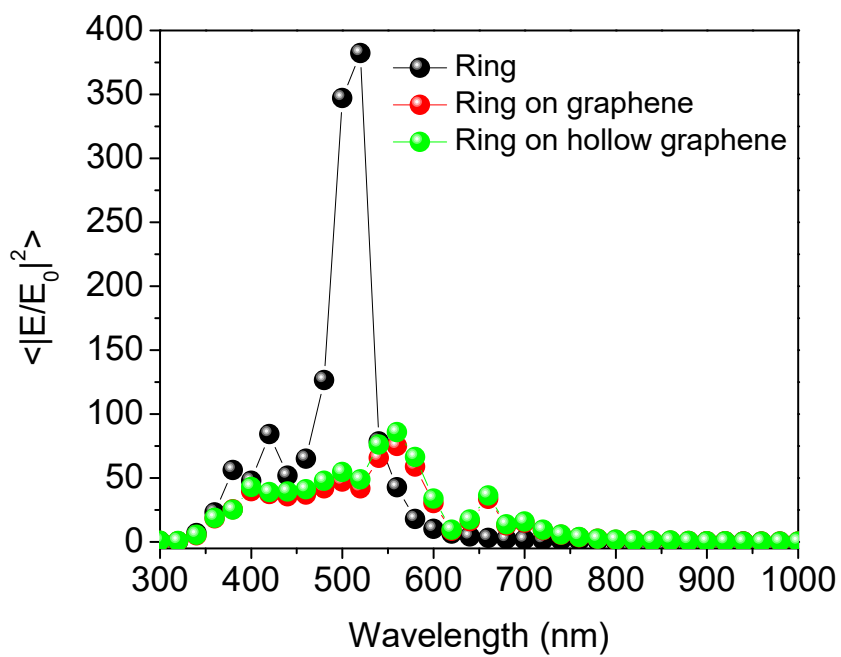

**Figure S13:** PRX-AgNR numerical simulations.

**SUPPORTING REFERENCES:**